\pdfoutput=1
\documentclass[runningheads]{llncs}
\usepackage{graphicx}
\graphicspath{{./figures/}}
\usepackage{color}
\usepackage{amsmath}
\usepackage[normalem]{ulem}
\DeclareMathOperator*{\argmaxC}{\arg\max} 
\usepackage{multirow}
\usepackage{booktabs}
\usepackage{subfig}
\usepackage{caption}
\usepackage{hyperref}
\begin{document}
\setlength{\abovedisplayskip}{2pt}
\setlength{\belowdisplayskip}{2pt}
\title{Bayesian Optimization on Large Graphs via a \\ Graph Convolutional Generative Model: Application in Cardiac Model Personalization \vspace{-.4cm}}
\titlerunning{Bayesian Optimization via a Graph Convolutional Generative Model}
%
\author{Jwala Dhamala\inst{1} \and
Sandesh Ghimire\inst{1} \and
John L. Sapp\inst{2} \and \\
B. Milan Hor\'{a}\v{c}ek\inst{2} \and
Linwei Wang\inst{1}}
%
\authorrunning{J. Dhamala et al.}
%
\institute{Rochester Institute of Technology, New York, USA \and
Dalhousie University, Halifax, Canada}
%
\maketitle              
\vspace{-.5cm}
\begin{abstract}
Personalization of cardiac models involves the optimization of organ tissue properties that vary spatially over the non-Euclidean geometry model of the heart. To represent the high-dimensional (HD) unknown of tissue properties, most existing works rely on a low-dimensional (LD) partitioning of the geometrical model. While this exploits the geometry of the heart, it is of limited expressiveness to 
allow partitioning that is 
small enough for 
effective optimization. Recently, a variational auto-encoder (VAE) was utilized as a more expressive generative model to embed the HD optimization into the LD latent space. 
Its Euclidean nature, however, neglects the rich geometrical information in the heart. In this paper, we present a novel graph convolutional VAE to allow generative modeling of non-Euclidean data, and utilize it to embed Bayesian optimization of large graphs into a small latent space. This approach bridges the gap of previous works by introducing an expressive generative model that is able to incorporate the knowledge of spatial proximity and hierarchical compositionality of the underlying geometry. It further allows transferring of the learned features across different geometries, which was not possible with a regular VAE. We demonstrate these benefits of the presented method in synthetic and real data experiments of estimating tissue excitability in a cardiac electrophysiological model.
\vspace{-.2cm}
\keywords{Graph convolution  \and Variational auto-encoder \and Bayesian optimization \and Model personalization.\vspace{-.2cm}}
\end{abstract}
\vspace{-.8cm}
\section{Introduction}
\vspace{-.2cm}
Personalized computer models of the heart have shown promise in various clinical tasks such as 
risk stratification~\cite{arevalo2016arrhythmia,Trayanova_NB18}
and predicting treatment response~\cite{sermesant2012patient}. With advances in medical imaging technologies,  personalization of high-resolution anatomical models is now feasible. By contrast, 
organ tissue properties 
vary spatially 
over the three-dimensional anatomical domain but have to be estimated from indirect, sparse and noisy data, 
resulting in a difficult ill-posed optimization problem with a high-dimensional (HD) unknown.

Numerous works have been presented in estimating these spatially-varying tissue properties in the form of high-dimensional model parameters. Most existing methods choose to represent spatially-varying tissue properties via a low-dimensional (LD) partitioning of the underlying geometrical model, either
as pre-defined segments~\cite{wong2012strain},
or iteratively optimized in a  coarse-to-fine fashion~\cite{dhamala2017spatially,sermesant2012patient}. This LD-to-HD definition
directly exploits the spatial proximity 
and hierarchical composition of 
the underlying geometry. However, 
it is of 
such limited expressiveness 
that 
the number of partitioning 
is either too low 
to faithfully 
represent high-resolution tissue properties, or too high to allow effective optimization.
In contrast,
recent work presented the use of a data-driven generative model of HD tissue properties, via a variational auto-encoder (VAE),
to embed the optimization into a LD latent space~\cite{dhamala2018high}. Being more expressive, 
this VAE-based generative model is able to represent high-resolution tissue properties with a latent code sufficiently small for effective optimization. However, as the regular VAE  is defined over Euclidean data, it does not take into account the valuable geometry information in the data, nor 
does it allow transferring among different geometry 
without first establishing  point-by-point correspondence.  

If we view organ tissue properties over a 3D geometrical model
as an image, convolutional neural networks (CNNs) are a natural choice to 
incorporate knowledge of the spatial proximity and hierarchical composition of the image ~\cite{bronstein2017geometric}. 
However, standard CNNs have been most  successful on data with an underlying Euclidean structure (\textit{i.e.}, image grids). 
Generalizing CNNs to non-Euclidean domains is an emerging area of research~\cite{bronstein2017geometric}, where significant efforts have been presented on addressing the challenges of defining convolution~\cite{fey2018splinecnn}, pooling, and up-sampling operations~\cite{ying2018hierarchical}. However, most developments to non-Euclidean CNNs are focused on supervised discriminative networks. To date, very limited work have been presented to enable generative modeling of non-Euclidean data.

In this paper, we 
present a novel VAE architecture that 
allows generative modeling of data over non-Euclidean domains, 
and utilize this generative model to embed Bayesian optimization of large graphs into a LD manifold. The presented approach bridges the gap of previous works by introducing an expressive generative model that is able to represent high-resolution tissue properties with a small latent code, while incorporating the geometrical knowledge in the data 
and being transferable across geometries.
We evaluate the presented method in synthetic and real-data experiments of estimating tissue excitability in a cardiac electrophysiological model, 
where we compare the expressiveness of the generative model and the accuracy of the subsequent parameter optimization with those obtained by using a linear reconstruction model based on principal component analysis (PCA) and a regular fully-connected VAE~
\cite{dhamala2018high}. We further demonstrate the feasibility of transferring the presented non-Euclidean VAE across patients. To our knowledge, this is the first introduction of a graph convolutional VAE and its use to 
enable Bayesian optimization over large graphs. Our code is available at \href{https://github.com/jwaladhamala/BO-GVAE}{https://github.com/jwaladhamala/BO-GVAE}.
%
\vspace{-.3cm}
\section{Background: Models of Cardiac Electrophysiology}
\label{sec:model}
\vspace{-.3cm}
\textbf{Cardiac Electrophysiological Model:}
Among various computational models of cardiac electrophysiology, phenomenological models 
are widely used in parameter personalization as they are able to express key macroscopic properties of cardiac excitation  with a small number of parameters~\cite{sermesant2012patient,dhamala2017spatially}. 
We thus adopt the phenomenological two-variable Aliev-Panfilov (AP) model~\cite{aliev1996simple}:
\begin{align}
    \begin{split}
		{\partial u}/{\partial t}  &= \nabla(\mathbf{D} \nabla u) - cu(u-\theta)(u-1) - uv,\\
		{\partial v}/{\partial t}  &= 	\varepsilon(u,v)(-v - cu(u - \theta - 1)),
	\label{eq:AP}
\end{split}
\end{align}
where $u$ is the normalized transmembrane potential, and $v$ is the recovery current. Parameter $\mathit{\varepsilon}$ controls the coupling between $u$ and $v$, $\mathbf{D}$ is the diffusion tensor, $\mathit{c}$ controls the repolarization, and $\mathit{\theta}$ controls tissue excitability. 
As $u$ is most sensitive to parameter $\theta$ in the AP model~(\ref{eq:AP})~\cite{dhamala2017spatially}, we focus on its estimation and use standard literature values for the remaining parameters~\cite{aliev1996simple}. Solving the AP model on a 3D discrete cardiac anatomy of meshfree nodes~\cite{wang2010physiological},
we obtain a 3D electrophysiological model of the heart that describes the temporal evoluation of 3D transmembrane potential $\mathbf{u}(t,\pmb{\theta})$.\\

\textbf{Measurement Model:}
$\mathbf{u}(t,\pmb{\theta})$ is measured on the body surface following the quasi-static electromagnetic theory,
solving which on a discrete subject-specific heart-thorax mesh gives a linear relationship between $\mathbf{u}(t,\pmb{\theta})$ and its surface potential measurement $\mathbf{y}(t)$ as: $\mathbf{y}(t) = \mathbf{H}\mathbf{u}(t,\pmb{\theta})$~\cite{wang2010physiological}.
\vspace{-.3cm}
\section{Personalizing HD Parameters on Unstructured Meshes}
\vspace{-.2cm} 
We seek parameter $\pmb{\theta}$ that minimizes the sum of squared errors between model output $M(\pmb{\theta}) = \mathbf{H}\mathbf{u}(t,\pmb{\theta})$ 
and patient's measurements $\mathbf{y}_d(t)$ as:
\begin{align}
\hat{\pmb{\theta}} = \argmaxC_{ \pmb{\theta}}\{{-\sum_t||\mathbf{y}_d(t) - M(\pmb{\theta})||^2}\}.
\label{eq:objfunc}
\end{align}
Directly solving~(\ref{eq:objfunc}) for the spatially-distributed $\pmb{\theta}$ is difficult~\cite{dhamala2017spatially}. %
Below we describe how we learn a LD-to-HD generation of $\pmb{\theta}$ that accounts for the 
underlying geometry,  
and embed the HD optimization into the expressive LD manifold.
\vspace{-.3cm}
\subsection{Graph Convolutional VAE}
\vspace{-.2cm}
We 
model the generation of spatially-distributed $\pmb{\theta}$ 
with a VAE~\cite{kingma2013auto}. A VAE consists 
a probabilistic encoder network with parameters $\pmb{\alpha}$ that approximates the intractable true posterior density as $q_{\pmb{\alpha}}(\textbf{z}|\pmb{\theta})$; and a probabilistic decoder network with parameters $\pmb{\beta}$ that represents the likelihood 
as $p_{\pmb{\beta}}(\pmb{\theta}|\textbf{z})$. For data defined on a Euclidean grid, structural information is incorporated in VAE through CNNs. Here, data resides over the 3D heart geometry, on which standard convolution and pooling operations are not applicable. Below we present a novel VAE architecture that enables convolution, pooling, and unpooling over non-Euclidean geometry of the heart.\\

\vspace{-.4cm}
\textbf{Local Connectivity \& Graph Convolution:} 
We model the cardiac mesh as a graph: $\mathcal{G} = (\mathcal{V}, \mathcal{E}, \mathbf{U})$, where vertices $\mathcal{V}$ consist of all $N$ meshfree nodes and edges $\mathcal{E}$ exist between each meshfree node and its $k$ nearest neighbors. $\mathbf{U} \in [0,1]^{N \times N \times 3}$ consists of edge attributes $\pmb{\upsilon}(i,j)$, calculated as  the normalized  $(x_1-x_2,y_1-y_2,z_1-z_2)$ if an edge $(i,j) \in \mathcal{E}$ exists between vertices at $(x_1,y_1,z_1)$ and $(x_2,y_2,z_2)$ and 0 otherwise. On this graph, we use a convolution operator based on spatial continuous convolution kernels because it was shown to allow better generalization to similar graphs~\cite{fey2018splinecnn}. In specific, given the graph $\mathcal{G}$ 
and $M$-dimensional input features $\{\mathbf{f}(i) | i \in \mathcal{V}\}$, 
the $l$-th convolution kernel is:
\begin{align}
g_l(\pmb{\upsilon}) = \sum_{\mathbf{p} \in \mathcal{P}} w_{\mathbf{p},l} \prod_{i=1}^{d} N_{i,p_i}^{m}(\upsilon_i),
\end{align} 
where $\small{((N_{1,i}^{m})_{1\leq i \leq k_1}, \dots, (N_{d,i}^{m})_{1\leq i \leq k_d})}$ denotes $d$ open B-spline bases of degree $m$ based on equidistant knot vectors with $d$-dimensional kernel size of $\small{\mathbf{k} = (k_1, \dots, k_d)}$, $\small{\mathcal{P} = (N_{1,i}^{m})_i \times \dots \times N(_{d,i}^{m})}$ is the Cartesian product of the B-spline bases, and $w_{\mathbf{p},l}$ are the trainable parameters. Given kernel functions $\mathbf{g} = (g_1, \dots, g_M)$ and input features $\mathbf{f} \in \mathcal{R}^{M}$, the spatial convolution operator for each vertex $i\in \mathcal{V}$ with a neighborhood $\mathcal{N}(i)$ based on its edge connectivity is then defined as: 
\begin{align}
(\mathbf{f} * \mathbf{g}) (i) = \frac{1}{|\mathcal{N}(i)|}\sum_{l=1}^{M}\sum_{j \in \mathcal{N}(i)}  f_l(j) g_l(\pmb{\upsilon}(i,j)).
\end{align}
\begin{figure*}[!t]
	\centering
    \subfloat{\includegraphics[width=0.9\textwidth]{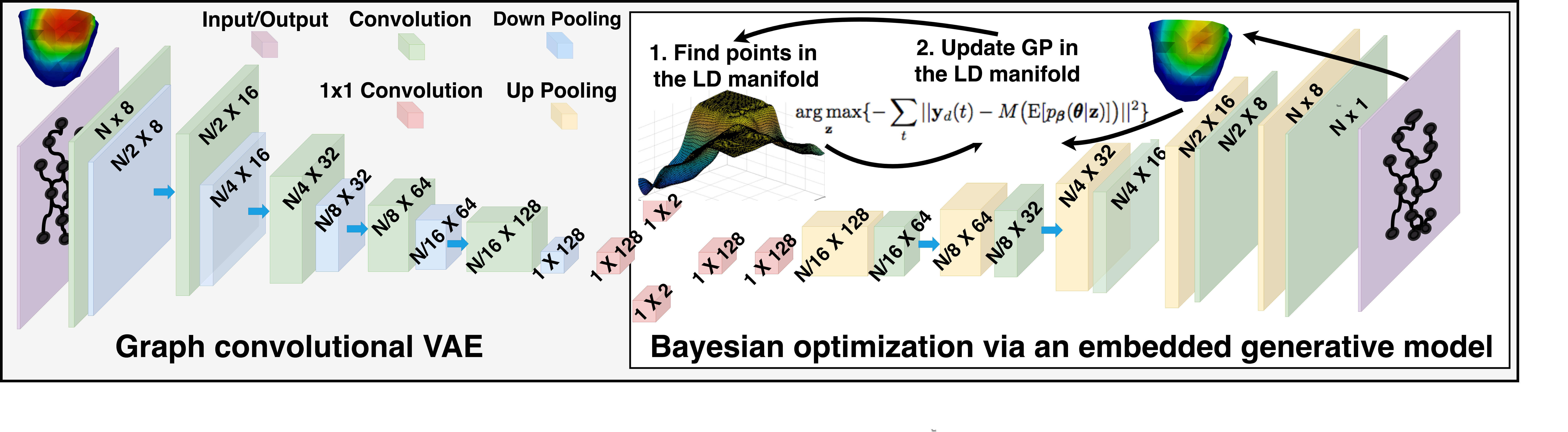}\vspace{-.5cm}}
     \caption{\small{Outline of the presented method, with dimensions labeled within the VAE.\vspace{-.5cm}}}
    \label{fig:net}
\end{figure*}

\textbf{Hierarchical Composition \& Pooling:}
To define pooling and unpooling operations necessary for the encoding-decoding architecture, a hierarchical representation of the graph is needed. We obtain this by an efficient multilevel graph clustering method  
based on minimizing the normalized cuts (Graclus)~\cite{dhillon2007weighted}, which reduces the graph size by half the number of vertices at each coarsening. 

We store hierarchical graph representation in matrices which reduces pooling/unpooling operations to efficient matrix multiplications. In specific, if $\mathcal{G}$ is a graph with $N_1$ vertices and $\mathcal{G}_c$ is its coarsened graph with $N_2<N_1$ vertices, we populate a binary matrix $\mathbf{P}
^{N_1 \times N_2}$, where $\mathbf{P}_{ij}=1$ if the $i^{th}$ vertex in $\mathcal{G}$ was grouped to the $j^{th}$ vertex in $\mathcal{G}_c$ and $\mathbf{P}_{ij}=0$ otherwise.  
Given $M$ feature maps
$\mathbf{F} \in \mathcal{R}^{N_1 \times M}$  over the vertices of graph $\mathcal{G}$ and $\mathbf{F}_c \in \mathcal{R}^{N_2 \times M}$ over graph $\mathcal{G}_c$, the average pooling in the encoder can be obtained by $\mathbf{F}_c = \mathbf{P}_{n}^\mathrm{T}\mathbf{F}$ and unpooling in the decoder 
by $\mathbf{F} =\mathbf{P}\mathbf{F}_c$, where $\mathbf{P}_{n}$ is column normalized from $\mathbf{P}$. 

\textbf{Graph Convolutional VAE:}
Using these building blocks along with global pooling and standard convolution layers, 
we construct 
a VAE architecture as shown in Fig.~\ref{fig:net}. 
It is trained by optimizing the variational lower bound on the marginal likelihood of 
the training data 
$\pmb{\Theta} = \{\pmb{\theta}^{(i)}\}_{i=1}^N$:
$\mathcal{L}(\pmb{\alpha};\pmb{\beta};\pmb{\theta}^{(i)}) = -D_{\mathrm{KL}} (q_{\pmb{\alpha}}(\textbf{z}|\pmb{\theta}^{(i)}) || p(\textbf{z})) + E_{q_{\alpha}(\textbf{z}|\pmb{\theta}^{(i)})} [\mathrm{log} p_{\pmb{\beta}}(\pmb{\theta}^{(i)}|\textbf{z})]
$. We set $q_{\pmb{\alpha}}(\textbf{z}|\pmb{\theta})$ and $p_{\pmb{\beta}}(\pmb{\theta}|\textbf{z})$ to be Gaussian parameterized by the graph convolutional networks. The prior $p(\textbf{z})\sim\mathcal{N}(0,1)$ is set to be an isotropic Gaussian, producing an analytical form for the KL divergence. 
Using the re-parameterization trick~\cite{kingma2013auto}, to express the random $\mathbf{z}$ as a deterministic variable, 
standard stochastic gradient methods can be used to optimize $\mathcal{L}(\pmb{\alpha};\pmb{\beta};\pmb{\theta}^{(i)})$.

\vspace{-.3cm}
\subsection{Bayesian Optimization on Large Graphs}
\vspace{-.2cm}
Bayesian optimization is a popular choice 
in optimizing complex 
objective functions such as (\ref{eq:objfunc}) 
~\cite{brochu2010tutorial}. It begins by defining a 
surrogate over the objective function. The optimization then consists of two iterative steps: 1) actively find a point that optimizes a utility function based on the surrogate, 
and 2) update the surrogate with the newly-selected point. Direct Bayesian optimization over HD space is difficult and its use over large graphs has not been reported~\cite{brochu2010tutorial,kandasamy2015high}. To enable this, 
we reformulate the original objective function in (\ref{eq:objfunc}) 
by replacing the unknown $\pmb{\theta}$ with the expectation of its non-Euclidean generative model:
\begin{align}
\hat{\mathbf{z}} = \argmaxC_{ \mathbf{z}}\{{-\sum_t||\mathbf{y}_d(t) - M\big(\mathrm{E}[p_{\pmb{\beta}}(\pmb{\theta}|\textbf{z})]\big)||^2}\}.
\label{eq:objfunc_new}
\end{align}
This allows us to embed surrogate construction and active selection of training points in a LD manifold. 
We initialize the Gaussian process (GP) surrogate of~(\ref{eq:objfunc_new}) with a zero mean function and an anisotropic M\'{a}tern 5/2 kernel.

\textbf{Active Selection of Training Points:}
To select a training point, we maximize the expected improvement (EI) utility function that favours a point with the highest expected improvement over the current optimum $f^+$~\cite{brochu2010tutorial,shahriari2016taking}:   
\begin{align}
\mathrm{EI(}\mathbf{z})& = (\mu(\mathbf{z})-f^+)\mathrm{\Phi}\bigg(\frac{\mu(\mathbf{z})-f^\textrm{+}}{\sigma(\mathbf{z})}\bigg) + \sigma(\mathbf{z}){\phi}\bigg(\frac{\mu(\mathbf{z})-f^\textrm{+}}{\sigma(\mathbf{z})}\bigg),
\label{eq:ei}
\end{align}
where 
$\mu(\mathbf{z})$ and $\sigma(\mathbf{z})$ are the predictive mean and standard deviation of the GP,  $\Phi$ is the cumulative normal  distribution, and $\phi$ is the normal density function. The first term promotes exploitation by favoring regions of high $\mu(\mathbf{z})$, while the second term promotes exploration by favoring the regions with high $\sigma(\mathbf{z})$.

\textbf{GP Update:} After picking a new point $\mathbf{z}^{(i)}$, the value of the optimization objective (\ref{eq:objfunc_new}) is evaluated at $\mathbf{z}^{(i)}$ as $\mathcal{J}^{(i)}$. The GP is then updated by including the new input-output pair of $(\mathbf{z}^{(i)},\mathcal{J}^{(i)})$ and maximizing the log marginal likelihood for kernel hyperparameters: length-scales and co-variance amplitude.
\vspace{-.3cm}
\section{Synthetic experiments}
\label{sec:syn_exp}
\vspace{-.3cm}
We evaluate the presented graph convolutional VAE (termed as gVAE) by: 1) its reconstruction accuracy
, and 2) optimization accuracy of the gVAE-based Bayesian optimization, both in comparison to existing methods. Accuracy is evaluated by the sum of squared error (SSE) in 
$\pmb{\theta}$, and the dice coefficient (DC) of the abnormal region 
obtained by thresholding $\pmb{\theta}$ with Otsu's method~\cite{otsu1975threshold}.%

We synthetically generate data of heterogeneous tissue excitability via random region growing in each cardiac model. Beginning with a single meshfree node as abnormal, we randomly grow the abnormal region 
by adding one of the nearest neighbors of the abnormal nodes, until the abnormal region 
reaches a desired size ($2\%$ to $40\%$ of the heart volume). 
On average, we generated $78,208\pm12,541$ data for training and $13,545\pm7654$ data each for validation and testing. The various layers and sizes of feature maps in each layer of the presented gVAE architecture are detailed in Fig.~\ref{fig:net}. 
We use B-spline basis degree of $m=1$ with kernel size of $k_1=k_2=k_3= 5$ in all graph convolution layers. 
All models are trained with a learning rate of $0.001$ with Adam optimizer~\cite{kingma2014adam}.
\begin{table*}[!t]
\small
\label{tbl:compare_nn}
\centering
\scalebox{0.8}{%
\begin{tabular}{@{}lccccc|ccccc|c@{}}
\toprule
\multicolumn{1}{c}{}                 & \multicolumn{5}{c|}{\textbf{SSE}}                                       & \multicolumn{5}{c|}{\textbf{DC}}                                               & \multirow{2}{*}{\textbf{\begin{tabular}[c]{@{}c@{}} Trainable\\ parameters \end{tabular}}} \\ \cmidrule(r){1-11}
\multicolumn{1}{c}{\textbf{Anatomy}} & \textbf{1} & \textbf{2}    & \textbf{3}    & \textbf{4} & \textbf{5}     & \textbf{1}     & \textbf{2}     & \textbf{3}     & \textbf{4} & \textbf{5}     &                                                                                              \\ \midrule
\textbf{PCA}                         & 12.73      & 14.18         & 12.35         & 23.85      & 23.26          & 39.80          & 39.87          & 41.31          & 49.17      & 54.42          & NA                                                                                           \\
\textbf{fVAE-3h}~\cite{dhamala2018high}                        & 8.03       & 8.45          & 8.44          & 13.07      & 13.70          & 61.77          & 66.20          & 60.45          & 70.30      & 70.72          & 2,087,822                                                                                    \\
\textbf{fVAE-4h}                      & 7.97       & 8.29          & 8.33          & 12.47      & 13.99          & 61.76          & 64.60          & 61.58          & 71.51      & 69.84          & 2,613,134                                                                                    \\
\textbf{fVAE-5h}                      & 7.42       & 8.01          & 8.19          & 13.96      & 12.41          & 64.59          & 65.72          & 62.21          & 68.04      & 74.50          & 3,138,446                                                                                    \\
\textbf{gVAE}                        & \textbf{6.66}       & \textbf{6.89} & \textbf{6.79} & \textbf{11.28}      & \textbf{11.43} & \textbf{68.43} & \textbf{70.92} & \textbf{70.70} & \textbf{75.10}      & \textbf{76.86} & 2,778,069                                                                                   \\ \bottomrule
\end{tabular}
}
\caption{\small{Comparison of reconstruction accuracy with the presented gVAE, PCA, and fVAE with various depths~\cite{dhamala2018high} in five different geometries.}\vspace{-.6cm}}
\end{table*}
\begin{figure*}[t]
	\centering
    \subfloat{\includegraphics[width=0.85\textwidth]{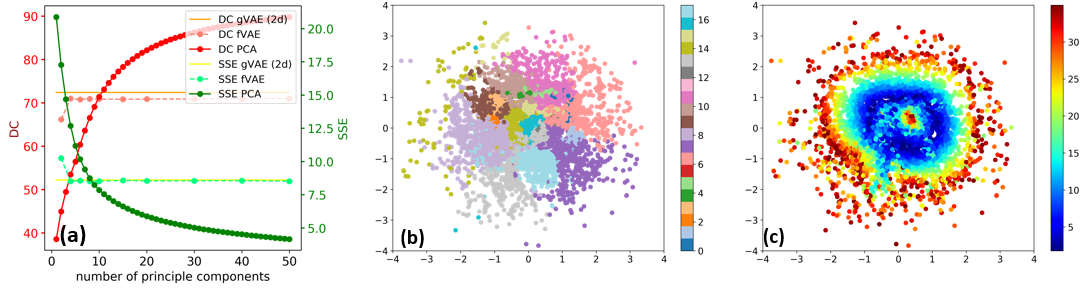}}	    \vspace{-.3cm}
     \caption{\small{(a) Comparison of reconstruction accuracy using gVAE with a 2d manifold \emph{vs.} PCA and fVAE~\cite{dhamala2018high} with various-dimensional manifolds. (b)-(c): Plots of 2d latent codes from gVAE colored by infarct location (b) and infarct size (c).}\vspace{-.4cm}}
    \label{fig:latent}
\end{figure*}

\textbf{gVAE as a Generative Model:}
On five different human heart models constructed from CT images, we first evaluate the ability of the presented gVAE model to reconstruct tissue excitability in comparison to: 1) PCA-based linear reconstruction, and 2) fully-connected VAE (termed as fVAE)~\cite{dhamala2018high} with three to five hidden layers 
(termed as fVAE-3h, -4h, and -5h, respectively). As summarized in Table 1, PCA being a linear model has the lowest accuracy. Compared to fVAE, the reconstruction accuracy of 
gVAE is consistently higher in 
both DC and SSE, even when its number of trainable parameters is similar to or lower than fVAE. However, gVAE is more expensive to train: 37.21 hrs \emph{vs.} 7.51 mins for fVAE-3h in TITAN Xp GPU. Nevertheless, note that 
the number of trainable parameters for gVAE does not increase for larger meshes. 

We further compare 
gVAE with two-dimensional (2d) latent codes \emph{vs.} fVAE and PCA with various latent dimensions. As shown in Fig.~\ref{fig:latent}(a), to achieve the reconstruction accuracy of gVAE with 2d latent codes, at least 13 principles components are required with PCA: this increase in 
dimension will make the subsequent optimization difficult. 
With fVAE, a similar SSE is attained with four latent dimensions, while a similar DC could not be attained with even 50 latent dimensions. This may be because fVAE does not consider the geometry underlying the spatial distribution of tissue excitability. Note that this increase in the number of latent dimensions will increase both the difficulty and the computational cost of the subsequent optimization. Fig.~\ref{fig:latent}(b) and (c) show that the latent code learned with gVAE are clustered by the location of the abnormal tissue and its radial direction encodes the size of the abnormal tissue.

\textbf{gVAE-based Parameter Optimization:}
On 40 synthetic cases on two different heart geometries, we conduct experiments on estimating unknown tissue excitability. In each case, we set an abnormal region by using various combinations of AHA segments 
which are very different from
the training set. Measurement data was simulated, with models described in Section~\ref{sec:model}, and then 
sub-sampled and corrupted with 20dB Gaussian noise. We compare the accuracy of gVAE-based Bayesian optimization with three previous approaches: 1) optimization on 17 fixed segments (termed as FS)~\cite{wong2012strain,sermesant2012patient}, 2) coarse-to-fine optimization along a fixed multi-scale mesh hierarchy (termed as FH)~\cite{dhamala2017spatially}, and 3) optimization on a LD manifold obtained with fVAE-3h~\cite{dhamala2018high}.  
\begin{figure*}[!t]
	\centering
    \subfloat{\includegraphics[width=0.9\textwidth]{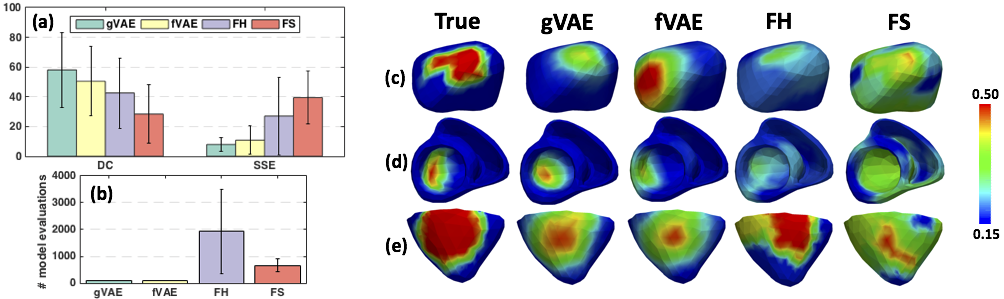}}\vspace{-.3cm}
     \caption{\small{(a) Accuracy and (b) computational cost of gVAE (green) versus fVAE~\cite{dhamala2018high} (yellow), FH~\cite{dhamala2017spatially} (purple), and FS~\cite{wong2012strain} (red). (c)-(e): Examples of estimated parameters.}\vspace{-.3cm}}
    \label{fig:syn_stat}
\end{figure*}
Result in Fig.~\ref{fig:syn_stat}(a) shows that gVAE-based Bayesian optimization is more accurate than all other approaches in both DC and SSE (paired t-test, $p < 0.05$). This indicates that gVAE was better at capturing the LD generative factors necessary for accurate optimization. Fig.~\ref{fig:syn_stat}(c)-(f) shows visual comparison on a few examples. The computational cost of gVAE-based optimization is much lower than that of FH ($>22$x) and FS ($>7.5$x), and similar to that of fVAE-based optimization.
\begin{figure*}[!t]
	\centering
    \subfloat{\includegraphics[width=0.95\textwidth]{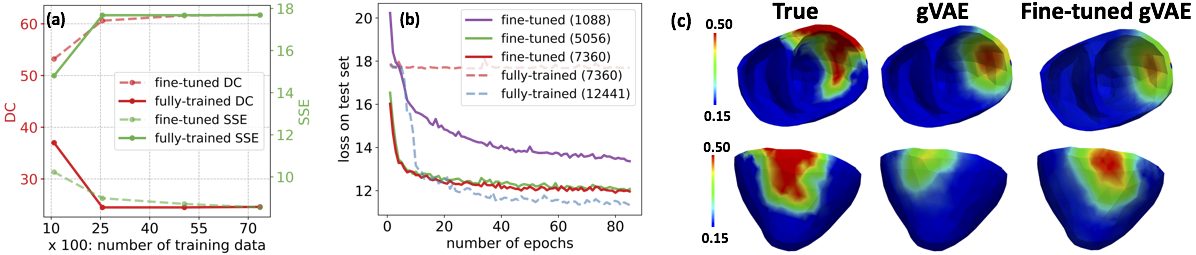}}\vspace{-.3cm}	    
     \caption{\small{(a) Reconstruction accuracy and (b) convergence of test losses from fine-tuned gVAE \emph{vs.} gVAE trained form scratch (termed fully-trained) with varying training data size. 
     (c) Examples of estimated parameters using gVAE fine-tuned with 7360 data.}\vspace{-.3cm}}
    \label{fig:transfer}
\end{figure*}

\textbf{Feature Sharing across Geometries:}
To demonstrate the feasibility of transferring the presented gVAE across different geometries, we take a pre-trained gVAE, fix the learned features in the encoder's graph convolution layers, and fine-tune the remaining layers for a different anatomy. We compare this training strategy to training a gVAE from scratch. Results in Fig.~\ref{fig:transfer}(a) show that a pre-trained model can be fined-tuned with as small as 1088 new examples.  
In comparison, gVAE could not be trained from scratch with $\leq 7360$ samples, shown both by the low reconstruction accuracy (Fig.~\ref{fig:transfer}(a): DC = 0.24; SSE = 17.68) and a flat test loss plot (Fig.~\ref{fig:transfer}(b)). Test loss plots in Fig.~\ref{fig:transfer}(b) also show that a pre-trained model starts with a lower loss and a larger size of training data leads to faster convergence. Parameter optimization via a gVAE fine-tuned with 7360 data on 20 cases achieved an average DC and SSE of 53.10 and 11.01 respectively. Fig.~\ref{fig:transfer}(c) shows some examples of the estimated parameters. More detailed analysis and experimentation is left to future work.

\vspace{-.3cm}
\section{Real Data Experiments}
\begin{figure*}[!t]
	\centering
    \subfloat{\includegraphics[width=0.85\textwidth]{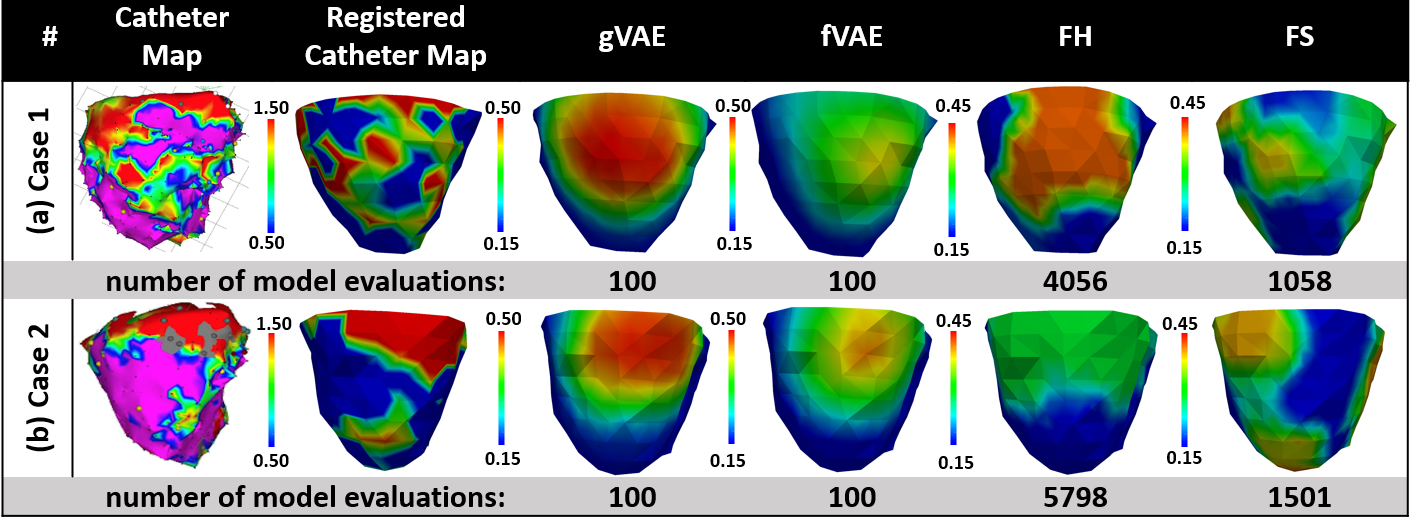}}	  \vspace{-.3cm}
     \caption{\small{Estimated parameters with gVAE, fVAE, FH, and FS on real-data studies.}\vspace{-.4cm}}
    \label{fig:real_exp}
\end{figure*}

\vspace{-.3cm}
We conduct real-data studies on two patients who underwent catheter ablation of ventricular tachycardia due to chronic myocardial infarction. Patient-specific heart-thorax models are obtained from axial CT images. Using 120-lead ECG as measurements, we evaluate the performance of 
the presented gVAE in estimating tissue excitability in comparison to the fVAE~\cite{dhamala2018high}, FH~\cite{dhamala2017spatially}, and FS~\cite{wong2012strain} methods. Training dataset and network architectures are as described in Section~\ref{sec:syn_exp}. We qualitatively evaluate the results 
with \emph{\textit{in-vivo}} catheter mapping data which, as shown in 
Fig.~\ref{fig:real_exp}, provides a reference for the location of the abnormal (red, voltage $\le0.5\mathrm{mV}$) and healthy (purple, voltage $>1.5\mathrm{mV}$)  regions. 

Case 1 has a large  heterogeneous abnormal region in the lateral LV region (Fig.~\ref{fig:real_exp}(a)). All methods are able to localize this region, but with varying degree of heterogeneity. Optimizations based on gVAE and fVAE are much faster requiring only 100 model evaluations, in comparison to FH and FS that required 4056 and 1058 model evaluations, respectively. By contrast, case 2 has a smaller but dense abnormal region in the lateral LV (Fig.~\ref{fig:real_exp}(b)). While all methods identify the general location of this abnormality, gVAE more accurately differentiates the region of dense core and border. In comparison, fVAE and FH estimate a larger border region; and the abnormal region revealed by FS is less accurate. Again, in this case, gVAE and fVAE required only 100 model evaluations, whereas FH and FS required 5798 and 1501 model evaluations, respectively.
\vspace{-.3cm}
\section{Conclusion}
\vspace{-.3cm}
We presented a novel graph convolutional VAE model that allows generative modeling of data defined over non-Euclidean structures and integrated it with the Bayesian optimization to enable optimization on large graphs. Experiments on estimating tissue properties distributed on a 3D cardiac mesh shows higher accuracy in terms of both reconstructing the tissue excitability and estimating them from indirect measurements, both in comparison to existing baselines. In future, we will incorporate realistic data from high resolution 3D images and investigate efficient transfer learning methods for models trained on such data.
\vspace{-.3cm}
\bibliographystyle{splncs04}
\bibliography{main}

\begin{thebibliography}{10}
\providecommand{\url}[1]{\texttt{#1}}
\providecommand{\urlprefix}{URL }
\providecommand{\doi}[1]{https://doi.org/#1}

\bibitem{aliev1996simple}
Aliev, R.R., Panfilov, A.V.: A simple two-variable model of cardiac excitation.
  Chaos, Solitons \& Fractals  \textbf{7}(3),  293--301 (1996)

\bibitem{arevalo2016arrhythmia}
Arevalo, H.J., Vadakkumpadan, F., Guallar, E., Jebb, A., Malamas, P., Wu, K.C.,
  Trayanova, N.A.: Arrhythmia risk stratification of patients after myocardial
  infarction using personalized heart models. Nature communications  \textbf{7}
  (2016)

\bibitem{brochu2010tutorial}
Brochu, E., Cora, V.M., De~Freitas, N.: A tutorial on bayesian optimization of
  expensive cost functions, with application to active user modeling and
  hierarchical reinforcement learning. arXiv preprint arXiv:1012.2599  (2010)

\bibitem{bronstein2017geometric}
Bronstein, M.M., Bruna, J., LeCun, Y., et~al.: Geometric deep learning: going
  beyond euclidean data. IEEE Signal Processing Magazine  \textbf{34}(4),
  18--42 (2017)

\bibitem{dhamala2017spatially}
Dhamala, J., Arevalo, H., Sapp, J., et~al.: Spatially-adaptive multi-scale
  optimization for local parameter estimation in cardiac electrophysiology.
  IEEE TMI  (2017)

\bibitem{dhamala2018high}
Dhamala, J., Ghimire, S., Sapp, J.L., Hor{\'a}{\v{c}}ek, B.M., Wang, L.:
  High-dimensional bayesian optimization of personalized cardiac model
  parameters via an embedded generative model. In: MICCAI. pp. 499--507.
  Springer (2018)

\bibitem{dhillon2007weighted}
Dhillon, I.S., Guan, Y., Kulis, B.: Weighted graph cuts without eigenvectors a
  multilevel approach. IEEE TPAMI  \textbf{29}(11),  1944--1957 (2007)

\bibitem{fey2018splinecnn}
Fey, M., Eric~Lenssen, J., Weichert, F., M{\"u}ller, H.: Splinecnn: Fast
  geometric deep learning with continuous b-spline kernels. In: CVPR. pp.
  869--877 (2018)

\bibitem{kandasamy2015high}
Kandasamy, K., Schneider, J., P{\'o}czos, B.: High dimensional bayesian
  optimisation and bandits via additive models. In: International Conference on
  Machine Learning. pp. 295--304 (2015)

\bibitem{kingma2014adam}
Kingma, D.P., Ba, J.: Adam: A method for stochastic optimization. arXiv
  preprint arXiv:1412.6980  (2014)

\bibitem{kingma2013auto}
Kingma, D.P., Welling, M.: Auto-encoding variational bayes. arXiv preprint
  arXiv:1312.6114  (2013)

\bibitem{otsu1975threshold}
Otsu, N.: A threshold selection method from gray-level histograms. Automatica
  \textbf{11}(285-296),  23--27 (1975)

\bibitem{Trayanova_NB18}
Prakosa, A., Arevalo, H.J., Deng, D., Boyle, P.M., Nikolov, P.P., Ashikaga, H.,
  et~al.: Personalized virtual-heart technology for guiding the ablation of
  infarct-related ventricular tachycardia. Nature Biomedical Engineering
  (2018)

\bibitem{sermesant2012patient}
Sermesant, M., Chabiniok, R., Chinchapatnam, P., et~al.: Patient-specific
  electromechanical models of the heart for the prediction of pacing acute
  effects in crt: a preliminary clinical validation. Medical image analysis
  \textbf{16}(1),  201--215 (2012)

\bibitem{shahriari2016taking}
Shahriari, B., Swersky, K., Wang, Z., Adams, R.P., De~Freitas, N.: Taking the
  human out of the loop: A review of bayesian optimization. Proceedings of the
  IEEE  \textbf{104}(1),  148--175 (2016)

\bibitem{wang2010physiological}
Wang, L., Zhang, H., Wong, K.C., Liu, H., Shi, P.:
  Physiological-model-constrained noninvasive reconstruction of volumetric
  myocardial transmembrane potentials. IEEE Trans. on Biomed. Eng.
  \textbf{57}(2),  296--315 (2010)

\bibitem{wong2012strain}
Wong, K.C., Relan, J., Wang, L., Sermesant, M., Delingette, H., Ayache, N.,
  Shi, P.: Strain-based regional nonlinear cardiac material properties
  estimation from medical images. In: MICCAI. pp. 617--624. Springer (2012)

\bibitem{ying2018hierarchical}
Ying, Z., You, J., Morris, C., Ren, X., Leskovec, J.: Hierarchical graph
  representation learning with differentiable pooling. In: NeurIPS. pp.
  4805--4815 (2018)

\end{thebibliography}
\end{document}